\documentclass[format=acmsmall]{acmart}

\AtBeginDocument{%
  \providecommand\BibTeX{{%
    \normalfont B\kern-0.5em{\scshape i\kern-0.25em b}\kern-0.8em\TeX}}}

\setcopyright{acmcopyright}
\acmJournal{PACMHCI}
\acmYear{2022}
\acmVolume{6}
\acmNumber{GROUP}
\acmDOI{10.1145/XXXXX}

\begin{document}

\title[Why Talk About Bias When We Mean Power?]{Studying Up Machine Learning Data: Why Talk About Bias When We Mean Power?}

\author{Milagros Miceli}
\affiliation{%
  \institution{Technische Universit{\"a}t Berlin, Weizenbaum Institute}
  \city{Berlin}
  \country{Germany}
}

\author{Julian Posada}
\affiliation{%
 \institution{University of Toronto, Schwartz Reisman Institute}
 \city{Toronto}
 \country{Canada}
}

\author{Tianling Yang}
\affiliation{%
  \institution{Technische Universit{\"a}t Berlin, Weizenbaum Institute}
  \city{Berlin}
  \country{Germany}
}

\begin{abstract}

Research in machine learning (ML) has primarily argued that models trained on incomplete or biased datasets can lead to discriminatory outputs. In this commentary, we propose moving the research focus beyond bias-oriented framings by adopting a power-aware perspective to “study up” ML datasets. This means accounting for historical inequities, labor conditions, and epistemological standpoints inscribed in data. We draw on HCI and CSCW work to support our argument, critically analyze previous research, and point at two co-existing lines of work within our community \,---\,one bias-oriented, the other power-aware.  This way, we highlight the need for dialogue and cooperation in three areas: data quality, data work, and data documentation. In the first area, we argue that reducing societal problems to “bias” misses the context-based nature of data. In the second one, we highlight the corporate forces and market imperatives involved in the labor of data workers that subsequently shape ML datasets. Finally, we propose expanding current transparency-oriented efforts in dataset documentation to reflect the social contexts of data design and production.

\end{abstract}



\maketitle

\section{Introduction}

In 2015, Facebook’s “real name” policy caught some media attention after the platform’s algorithm failed to recognize the names of hundreds of North American Indigenous users as “real” and proceeded to cancel their accounts \cite{Sampat2015, haimson2016}.  According to Facebook’s algorithm, real names seemed to be defined by Anglo-Western conventions. Thus, the system flagged names composed of several words or with unusual capitalization. Moreover, despite the many contextual factors that determine how a name sounds and looks like, Facebook enforces its policy algorithmically, that is, in a narrow, unquestionable, and predefined way. 

At first sight, the issues raised by users whose names were flagged could indicate the presence of biased training data: As Anglo-Western names are dominant and those from other cultures are underrepresented, the unbalanced dataset leads to “unfairness.” This approach is not wrong, but it remains insufficient to fully address the issue at stake: that some worldviews are considered more valid than others. Framing this type of issue as “bias” tends to obscure a set of persistent questions behind and beyond the technical domain: What \emph{is} a real name? Who decides over the \emph{realness} of a name? And, do we need a real name policy at all?

In the past decade, injustice and harm produced by data-driven systems have often been addressed under the umbrella term “bias.” Research has shown that biases can penetrate ML systems at every layer of the pipeline, including data, design, model, and application \cite{olteanu2016}. Special attention has been paid to the quality of data, arguing that models trained on incomplete or biased datasets can lead to discriminatory or exclusionary outcomes \cite{olteanu2016,dixon2018,buolamwini2018}.  Moreover, significant academic focus lies upon bias in data work and crowdsourcing \cite{brodley1999, cheng2013, finin2010, hube2019, fan2020}.  Because of the interpretative character of tasks such as labeling, rating, and sorting data, abundant research has focused on the individual subjectivities of data workers to account for biases in data, investigating ways of mitigating them by constraining workers’ judgment.

With the present commentary, we aim to contribute to the discussion around data bias, data worker bias, and data documentation by broadening the field of inquiry: \emph{from bias research towards an investigation of power differentials that shape data.} As we will argue in the following sections, the study of biases locates the problem within technical systems, either data or algorithms, and obscures its root causes. Moreover, the very understanding of bias and debiasing is inscribed with values, interests, and power relations that inform what counts as bias and what does not, what problems debiasing initiatives address, and what goals they aim to achieve. Conversely, the power-oriented perspective looks into technical systems but focuses on larger organizational and social contexts. It investigates the relations that intervene in data and system production and aims to make visible power asymmetries that inscribe particular values and preferences in them. 




Computing has become so widely integrated into society, both influencing and being shaped by it, that a broader understanding of sociotechnical systems becomes key to addressing social concerns surrounding their development and deployment. In this sense, “debiasing” ML data is not sufficient to fully address the questions posed by “real-name” algorithms and other data-driven systems that are deeply ingrained in our everyday lives. Such approaches could be expanded by applying a relational view on the power dynamics and the economic imperatives that drive machine learning, i.e., considering that biases do not occur in a vacuum but are fundamentally entangled with naturalized ways of doing things within the organizations where datasets and models are developed. This requires an epistemological shift in terms of how to think of these problems, what questions to ask, and what methods to use. Such a shift can only be achieved through more dialogue between Computer Science and disciplines such as Sociology, Anthropology, and Economy. Given the important interdisciplinary tradition in HCI and CSCW, we believe in the key role of these communities in prompting conversations around power and ML systems. 

On that basis, we follow the line of previous work \cite{barabas2020, Forsythe2001, seaver2014} that has borrowed the concept of “studying up” from anthropologist Laura Nader. In anthropology, studying up means expanding the field of inquiry to study power, i.e., interrogating elites that have remained significantly understudied in the anthropological tradition. In their call to study up algorithmic fairness, Barabas and her colleagues \cite{barabas2020} explain that this endeavor “requires a new set of reflective practices which push the data scientist to examine the political economy of their research and their own positionality as researchers working in broken social systems.” In a similar vein, our appeal is to “study up” machine learning data by investigating labor conditions, institutional practices, infrastructures, and epistemological stances encoded into datasets, instead of looking for biases at the individual level and proposing purely technical interventions. 

In the following sections, we will zoom into three critical ML-related fields of inquiry: \emph{data quality, data work, and data documentation}.
While our argument is based on previous research, it is worth mentioning that a systematic literature review is not within the scope of this commentary. Here, we look into CSCW and HCI research to critically discuss work that revolves around the concept of “bias,” while building on previous initiatives that have, instead, striven for a more comprehensive understanding of sociotechnical systems.  By contrasting both perspectives that co-exist within our research communities, we highlight the importance of fostering more dialogue between them to produce research that expands the investigation of individual biases into a consideration of power asymmetries within organizations and among them. Our argumentation concludes with suggestions as to how to study up machine learning data and why.   

\section{The Limits of Bias}

Studies on data and algorithmic biases have demonstrated how data-driven systems can enhance discriminatory practices and result in exclusionary experiences in various domains, including credit \cite{lee2021a, daniellek.citron2014} and algorithmic filtering \cite{noble2018, baker2013}. CSCW and HCI research has explored algorithmic bias in the job market \cite{chen2018c, hannak2017}, advertisement \cite{ali2019}, and image search engines \cite{kay2015a}, among several other domains. Moreover, researchers have shown how algorithms contribute significantly to the visibility of information \cite{kulshrestha2017} and how stereotypes are perpetuated by gender recognition systems \cite{hamidi2018a}. The quest for addressing these problems has prompted the development of an area of research that emphasizes the issue of bias, and the values of fairness, accountability, and transparency in mitigating its effects \cite{dignazio2020}. The fact that research in the technical realms takes issue with social inequities and examines the harmful effects of technology is a significant step. However, work in Critical Data and Algorithmic Studies, as well as CSCW and HCI, has argued for a shift of perspective from individual cases and individual biases towards the comprehensive analysis of social practices and power relations involved in creating the systems that surround us \cite{kalluri2020, birhane2021a, denton2020, barabas2020, miceli2020b, scheuerman2020}. 

Technological development is sociotechnical in nature and data, as an abstraction \cite{dignazio2020,kitchin2014}, is not given but created through human discretion \cite{muller2019} and shaped by power dynamics \cite{miceli2020b}. By focusing on technical solutions for personal subjectivities, bias-oriented approaches are mostly unable to account for the social processes underway that comprise increasing surveillance and privacy intrusion to satisfy the insatiable need for more and diverse data \cite{couldry2019}, and the important shifts in labor that include the mobilization of largely precarized workforces to process data and make it “readable” for ML systems \cite{Posada2020a}. Through a power-aware lens, it is possible to interrogate why accurate, efficient, and seemingly “debiased” ML systems are still not \emph{good for everyone}.  For example, accurate facial recognition used for surveillance is dangerous in the hands of unscrupulous organizations or oppressive governments.
Debiasing efforts sometimes mitigate harm, but machine learning will inevitably perpetuate injustices if systems remain controlled by powerful organizations that follow their own agenda. In this context, attempts to address and mitigate biases appear as “a tiny technological bandage for a much larger problem” \cite{dignazio2020}. Research efforts that focus on designing “debiased” systems are not bad. However, the question stands: “debiased” according to whom and for whom? \cite{kalluri2020}. The bias-oriented approach provides only limited tools to explore this and other important questions.

Moreover, framing sociotechnical problems as bias constitutes what Powles and Nissenbaum call “a seductive diversion” \cite{powles2018}: On the one hand, we are told that biases can be fought and mitigated, and that data can be cleaned and systems debiased. On the other hand, it is argued that bias is not only a technical but also a societal issue; hence, biases are everywhere and nowhere. If society is biased, then biased AI cannot be avoided. This way, bias-oriented framings present a puzzle that keeps us continually busy because technical fixes are inadequate solutions to societal issues. We are always on the way to identifying and mitigating biases in an attempt to build debiased systems while knowing that the ideal of a debiased system can never be achieved. 

Still, considerable efforts, both within and beyond HCI and CSCW, are invested in technical tools to mitigate data biases, algorithmic biases, and workers’ biases in domains where interrogation and reflexivity could be more fruitful. This way, the bias puzzle distracts us from addressing fundamental questions about who owns data and systems, who are the data workers, whose worldviews are imposed onto them, whose biases we are trying to mitigate, and what kind of power datasets perpetuate. “It also denies us the possibility of asking: should we be building these systems at all?” \cite{powles2018}. These questions could shift the perspective because they interrogate privileged and naturalized worldviews encoded in data and systems that (re)produce the status quo. Consequently, such questions are more about power than they are about bias. 

In the following section, we will develop this argument further by going deeper into the discussion of problematic aspects of framing power differentials and injustice as “bias” in ML data, data work, and dataset documentation.

\subsection{Data is Always Biased}

Data bias has been defined as “a systematic distortion in the data” that can be measured by  “contrasting a working data sample with reference samples drawn from different sources or contexts.”\cite{olteanu2016}  This definition encodes an important premise: \emph{that there is an absolute truth value in data and that bias is a “distortion” from that value.} This key premise broadly motivates approaches to "debias" data and ML systems. However, we argue that the problem with this assumption is that data never represents an absolute truth. Data, just like truth, is the product of subjective and asymmetrical social relations \cite{dignazio2020}.

In their groundbreaking analysis of three commercial gender classifiers made available by Microsoft, Face++, and IBM, Buolamwini and Gebru \cite{buolamwini2018} show that darker-skinned women are up to 44 times more likely to be misclassified than lighter-skinned men. This work is often cited as a paradigmatic example of how data can contain biases as related to the underrepresentation of certain groups. Looking at this problem from a bias-oriented perspective, the solution seems straightforward: add more and diverse data to training datasets. However, as Gebru also points out in an interview, biased data is only part of the story: “[…] not just bias in the training data, but ethics in general \,---\, what’s okay to do, what’s okay not to do, the power dynamics of who has data, who doesn’t have data, who has access to certain kinds of models, and who doesn’t” \cite{hunter-syed2020}. The contextual issues that escape technical fixes also include: where is diverse data \,---\, the “missing faces” \,---\, harvested? Under which conditions? Who classifies them? Moreover, and considering that Black and Brown populations have historically been subject to surveillance, persecution, and police violence \cite{Browne2015, benjamin2019}, it is worth asking if improving facial-recognition systems so that they can properly “see” dark-skinned faces would further perpetuate such injustice.

Our point is that biased data is undoubtedly one issue to consider when it comes to discriminatory outcomes from machine learning systems, but so are social structures, the definition of social problems to be solved in computational terms, and the widespread assumption that algorithms are neutral where people are not. These factors, as well as data, are deeply political. Machine learning systems are fundamentally trained to cluster and classify data. When these classifications are value-laden and interest-informed, they result in imposing and promoting the particular set of interpretations and worldviews of some groups, which could reinforce injustice \cite{boyd2018}. In other words, ML systems have real effects on real people. Therefore, it is important to consider that their quality cannot be thought of only in terms of accuracy and performance. Some issues do not just get solved by throwing in more data and quantification does not always lead to better representation or less harm. In a broader sense, harms produced by ML systems manifest existing power asymmetries: they are about having the power to decide how systems will “see” and classify, what data is worth including, and whose data we can afford to ignore. Those harms are about the power to impose a hegemonic worldview over others possible.

Tracing the links to historical and ongoing asymmetries can be helpful to understand how data comes to be \cite{denton2020} and what kind of political work ML systems perform \cite{hanna2020}. This means, of course, acknowledging that the data that fuels machine learning is produced by humans and hence is laden with subjective judgments. Even so, discussions around human intervention on data ought to consider that the subjective forces that shape data and systems are not just about the personal biases of individual actors. Data is produced within organizations and through practices that “embody specific technical ideals and business values” \cite{passi2018} that also shape the subjectivities of data workers. We are for sure not the first ones to make this statement: Researchers in Human-Centered Data Science (HCDS) \cite{kogan2020,muller2019, zhang2020, passi2017, passi2018, muller2021} and Human-Centered Machine Learning (HCML) \cite{Chancellor2019} have explored data as a “human-influenced entity”\cite{muller2019}. A series of CSCW/HCI workshops on Data Science work practices \cite{muller2019b,muller2020a} has fostered interesting conversations on collaboration, meaning making, trust, craft, and power. This line of work has shown that narratives, preferences, and values related to larger socio-economic contexts are embedded in processes of data production \cite{paullada2020}. Practices such as the framing of real-world questions as computational problems \cite{passi2019, berendt2019}, the choice of training data and data-capturing measurement interfaces \cite{pine2015}, the establishment of taxonomies to label data \cite{miceli2020b}, and the selection of data features \cite{muller2019} as well as the design of data to be recognizable, tractable, and analyzable \cite{feinberg2017, muller2019}, all are decisions that are hardly ever made by individual choice and in a vacuum. Instead, they concern organizational structures and depend on what is possible in terms of time and budgets, and what is expected in terms of computational output and revenue plan.  

As the examples in the following section will show, despite the abundant CSCW and HCI initiatives that have argued that “datasets aren’t simply raw materials to feed algorithms, but are political interventions” \cite{crawford2019}, a considerable number of investigations within those research communities still comprise the assumption that data represents an absolute truth value and that bias is just a distortion that can be mitigated. The problem is that framing arbitrary representations in data as bias misses the political character of datasets: there is no neutral data and no apolitical standpoint from where we can call out bias \cite{crawford2019}. Datasets are always “a worldview” \cite{davis2020} and, as such, data always remains biased. 

\subsection{“Mitigating Worker Biases” Should Not Be the Goal}

Datasets are conditioned by the networked systems in which they are created, developed, and deployed. The examination of data provenance and the work practices involved in dataset production is essential to the investigation of subjectivities embedded in data-driven systems \cite{passi2017, muller2019, muller2021, miceli2020b}. In formal terms, data work for machine learning involves tasks such as the collection, curation, and cleaning of data, labeling and keywording, and, in the case of image data, it can also involve semantic segmentation (i.e., marking and separating the different objects contained in a picture) \cite{Casilli2019, Casilli2019b, Tubaro2019}. Outsourced data workers perform these tasks through digital labour platforms (crowdsourcing) or business process outsourcing companies (BPOs). In this regard, outsourced data work is part of the broader gig economy landscape, in the case of platforms \cite{Woodcock2020}, and other digital service BPOs, like those providing content moderation \cite{Roberts2019}. In both cases, these types of work are characterized for low- or piece-wages, limited-to-no labor protection, and high levels of control and surveillance.

The tasks that data workers perform are fundamentally about making sense of data \cite{muller2019,miceli2020b}, that is, about interpreting the information contained in each data point. Because of the subjective character of data-related tasks, bias-oriented research in this space has focused mainly on the individual subjectivities of workers, considering their judgments to be a significant source of biases and data quality errors \cite{brodley1999, cheng2013, hube2019, wauthier2011, ghai2020}. For example, abundant research considers labelers’ subjectivity in annotation tasks to be one of the main reasons for biased labels. The field of research directed towards the study of crowdworkers and crowdsourcing platforms \cite{brodley1999, cheng2013, finin2010, hube2019, fan2020} offers several examples of such an approach. Some of this work argues, for example, that data workers’ cognitive biases \cite{eickhoff2018}, their own preferences \cite{ramanath2013}, and political stances \cite{yano2010} can negatively affect data. Moreover, research has proposed methods to identify and monitor annotator bias within datasets \cite{geva2019, artstein2005, hube2019, wauthier2011}. In a paper presented at CHI 2019, Hube et al.\cite{hube2019} explore how crowdworkers annotate machine learning data and propose a framework for mitigating their biases. The authors argue that extreme personal opinions of workers can affect data labeling tasks and produce biased data, especially when the tasks involve opinion detection and sentiment analysis. Consequently, they add that “the ability to mitigate biased judgments from workers is crucial in reducing noisy labels and creating higher quality data.” 
This research follows the line of many of the work in crowdsourcing that rests on three premises: (1) that data should represent an absolute ground truth, and that bias is a deviation from that truth value, (2) that data workers have enough agency to interpret data according to their personal judgment and could, therefore, be prone to deviating from the predefined truth value that data should represent, (3) and that workers using their own subjectivity to interpret data is \emph{per se} detrimental to the quality of data. 
Quite often, such approaches to detect and mitigate workers’ bias do not consider that data workers constitute automation’s “last mile” \cite{gray2019}, that is, the bottom end of hierarchical labor structures, and  that they collect and label data within organizational structures and according to predefined truth values instructed to them by managers and clients. 

Socio-technical systems are complex in nature and this also includes the data work that fuels them.  Several issues framed by previous research as “worker bias” are actually manifestations of broader power asymmetries that fundamentally shape data: power asymmetries that are as trivial as being the boss in a tech company and have decision-making power, or being an underpaid crowdworker who risks being banned from the platform if they do not follow instructions.  We argue that research that focuses on the personal biases of workers and aims at mitigating them could benefit from an interrogation of power differentials, normalized preconceptions, and profit-oriented interests that shape labor conditions in data work.

Let us look at some examples from our on-going research project that focuses on data work in Latin America \cite{miceli2021b}. These examples should provide an idea of the identity of the workers whose biases research attempts to mitigate. In this case, they are located, as with many data workers, in Argentina and Venezuela. The Venezuelan economy is currently experiencing the highest levels of inflation in the world and many people look for work with crowdsourcing platforms because they offer a steady income, paid in US dollars. Melba, one of the crowdworkers interviewed by us, is a retired woman. Her monthly pension is the equivalent to USD\$1, which, as she puts it, is “not enough to buy half a dozen eggs; not enough to buy a piece of cheese or bread.” The payment she receives for doing data work is also meager by international standards. However, in a country experiencing hyperinflation, it allows her to supplement her income. In the case of Juan, another crowdworker from Venezuela, the income from the platform is comparable to what he would receive doing harvest work in the neighboring country, Colombia. However, doing data annotation allows him to stay in Venezuela with his family instead of migrating and being apart. In the case of Argentina, most of the data workers we interviewed live in the impoverished areas that surround Buenos Aires. Despite the meager salaries they receive for data collection and annotation tasks (the equivalent to US\$1.80 per hour), and the exhausting nature of the work they perform, all interviewees expressed being proud of their work. For many of them, doing data work means finally having a desk job and breaking with generations of unlicensed cleaning or construction work. Similarly, for many of the Venezuelan crowdworkers, having access to this type of work means avoiding extreme poverty and having a means to circumvent many of the difficulties present in their local labour market.

The cases described above are not extreme or marginal. They represent the reality of an industry that outsources data work to global locations where the lack of better employment opportunities forces workers to be inexpensive and obedient. A growing body of literature in CSCW and HCI has taken crowdworkers' perspective and pointed to the issues of underpayment \cite{hara2018}, crowdworkers' growing dependency on performing crowdsourcing tasks to make ends meet \cite{ross2010}, the use of parameters and processes (e.g. the rate of previously approved and paid tasks) to select and recruit crowdworkers \cite{barbosa2019a}, and the power asymmetries introduced by crowdsourcing platform design and inherent in the relations between service requesters and crowdworkers \cite{irani2013, salehi2015a, martin2014}. Ekbia and Nardi use the term \textit{heteromation} to characterize the shift in technology-mediated work and labor in which human intervention and action are indispensable for technical systems to function \cite{ekbia2014}. They argue that heteromated systems, like the platform Amazon Mechanical Turk, are the outcome of socioeconomic forces rather than of the essential attributes of humans and machines, as commonly assumed \cite{ekbia2014}. The authors not only scrutinize the asymmetrical labor relations in which crowdworkers are put at a significantly disadvantaged position, but also emphasize that crowdworkers are regarded as mere “functionaries” of an algorithmic system and are rendered invisible \cite{ekbia2014}. Apart from drawing attention to invisible labor and asymmetrical labor relations, a political economic perspective further highlights the profit-driven imperative of capital, the surveillance and social control enabled and reinforced by digital technologies, and the political nature of design choices and technologies that mediate work and labor \cite{ekbia2015, ekbia2016}. 
These studies are important examples within CSCW and HCI of how shifting researcher's gaze upwards to look into power dynamics can expose fundamentally different issues with sociotechnical systems. However, they unfortunately have not received enough attention from scholars in those very same research communities that investigate bias in data work.


Social and labor conditions affect the dependency of workers on data work, and that dependency has an effect on how datasets are produced, such as restricting workers' ability to raise questions about annotation instructions and tasks. Expanding the question of how heteromated labor affects crowdworkers, broader communities, and polities \cite{ekbia2014}, we propose also asking \emph{how power asymmetries in heteromation inform machine learning datasets and systems}. Starting from the assumption that such power imbalances are the problem, not just bias, leads to fundamentally different research questions and methods of inquiry. We believe that this perspective can significantly contribute to broadening research on data worker and crowdsourcing bias.  

\subsection{Data Documentation Beyond Bias Mitigation}

Several frameworks and tools to document machine learning datasets and models have been proposed and applied. 
Significant examples are the work of Bender and Friedman with the \textit{Data Statements for Natural Language Processing} \cite{bender2018}, Holland et al. with the \textit{Dataset Nutrition Label} \cite{holland2018}, and most prominently, Mitchell et al. with \textit{Model Cards for Model Reporting} \cite{mitchell2019}, and Gebru et al. with \textit{Datasheets for Datasets} \cite{gebru2020}. In these investigations, data bias appears as a core motivation for developing documentation frameworks. The authors argue that documentation can help “diagnose sources of bias” \cite{holland2018}, and has potential to “mitigate unwanted biases in machine learning systems” \cite{gebru2020}. In the present subsection, we would like to discuss two ways to complement these data documentation approaches. The first one is to consider expanding the documentation of dataset composition beyond merely listing dataset’s elements. The second one is to consider the complex and intricate relationship between dataset creators and dataset consumers. As we will argue, both considerations could allow us to expand this line of research and explore power relations in machine learning through a CSCW-informed perspective beyond bias-oriented framings.

First, we argue for the inclusion of further information beyond the proposed list of data “ingredients”. For instance, one of the questions in \textit{Datasheets for Datasets} asks “does the dataset identify any subpopulations?” (e.g. by race, age, or gender). This way of documenting dataset composition is key but it also brings along what we consider to be a valid question: \emph{Is this information sufficient in itself to explicate unjust outcomes?}  Disclosing whether a dataset includes racial categories and listing said categories “does not speak to the problem of such categories’ reductiveness, nor makes the assumptions behind race classifications embedded in datasets explicit” \cite{miceli2021a}. We believe that documentation can and should tell us more, for instance, about how data collectors and annotators have established the correspondence between data point and category. Moreover, “to impose order onto an undifferentiated mass, to ascribe phenomena to a category \,---\, that is, to name a thing \,---\, is in turn a means of reifying the existence of that category” \cite{crawford2019}, as Crawford and Paglen put it. Similarly, when the documentation of racial categories contained in a dataset is limited to listing them without further reflection, the risk exists that the documentation could contribute to the reification and naturalization of such categories.

Our second idea is to look deeper into the intricate relationship between data workers and requesters. In their investigation, Gebru et al. \cite{gebru2020} argue that \textit{Datasheets for Datasets} would improve communication between dataset creators and dataset consumers. The clear differentiation between dataset creators and consumers surely applies to large open datasets commonly used for benchmarking, such as ImageNet. However, such a clear separation does not correspond with the totality of machine learning datasets or even to most of the ML products that are created for commercial use. For instance, Feinberg \cite{feinberg2017} unveils "a multilayered set of interlocking design activities" in data infrastructure, collection and aggregation in data production. In many settings of data production, design activities and decisions are shaped, if not determined, by dataset consumers and other external stakeholders rather than data workers, which makes them co-designers of datasets. In such settings, the distinction between consumers and producers is more ambiguous. Previous work \cite{kazimzade2020, miceli2021a} has explored companies producing (or outsourcing the production of) tailor-made datasets to train their own ML models. These companies have particular requirements in mind and produce data specifically tailored to the ML product they aim to develop. Many of these organizations do outsource data collection and labelling but, even then, tasks are completed according to the specific instructions provided by model developers \,---\, whom Gebru et al. call “dataset consumers.” Once the labeled dataset is sent to the model developers, data is further cleaned and sometimes re-labeled. In a similar vein, Seidelin \cite{seidelin2020}, building on and extending Feinberg’s design perspective of data, situates data work and practices in organizational, cross-organizational, and multi-stakeholder contexts. Her research reveals that data work and data-based services are by nature collaborative and cooperative, and that the design and production of data are rather co-design processes. These perspectives challenge the clear separation between dataset producers and consumers and show that dataset consumers are also dataset co-creators.

With both ideas described above, we seek to expand previous work in data documentation beyond bias-related motivations. Merely listing the composition of a dataset without interrogating the origins of its categories might be sufficient if the aim of documentation is “mitigating unwanted biases”. However, it is not enough to unveil the political work those categories perform. Similarly, the stiff differentiation between producers and consumers could reinforce a similar logic as the studies on worker bias described in the previous section: The responsibility for data quality issues lies with data workers exclusively and requesters have no control over assumptions encoded in datasets because they are mere “consumers.” We argue that data transparency could be better explored by moving the focus away from the documentation of datasets’ technical features and biases, to highlighting the importance of documenting production contexts, aiming to make visible the dynamics of power and negotiation that shape datasets. 

Such an extended perspective could also help to explicate why, despite growing calls for more transparency in machine learning, data documentation practices are still limited in machine learning. Some factors to take into account are that requesters often regard the information that should be documented as corporate secrecy and that documentation is often perceived as an optional task, in some cases even as a burden, that is time-consuming and expensive \cite{miceli2021a}. Moreover, the lack of knowledge and training, be they technical or ethical, makes data workers less equipped to reflect on what should go into documentation \cite{kazimzade2020} and, even among informed workers, hierarchical managerial structures in BPOs and the risk of being banned in data work platforms would probably make workers reluctant to use documentation to reflect upon taken-for-granted practices.  
To address such difficulties, researchers developing documentation frameworks could benefit from the acknowledgement that data production is a collaborative project which demands cooperative efforts from actors that hold different (organizational and social) positions and decision-making power to shape data \cite{miceli2020b, Dafoe2021}. While the bias-orientation of existing frameworks counteracts documentation’s potential to make power explicit and contestable, we believe that CSCW research could significantly contribute to this line of work.    
More than diagnosing “the source of bias,” documentation should aim at interrogating work practices and decision-making hierarchies within and among organizations. 


\section{Conclusion}

This commentary has critically explored several implications of framing diverse socio-technical problems as “bias” in machine learning. Through examples related to the study of ML datasets, data work, and dataset documentation, we have argued for a shift of perspective to orient efforts towards considering the effects of power asymmetries on data and systems. 

Such reorientation not only concerns privileged groups among machine learning practitioners. It is also about the role of researchers and the intertwined discourses in industry and academia \cite{green2020a}.
We need more research that interrogates the relationship between human subjectivities and (inter-)organizational structures in processes of data production. Most importantly, power-oriented investigations could allow researchers to “shift the gaze upward” \cite{barabas2020} and move beyond a simplistic view of individual behaviors and interpretations that, in many cases, ends up allocating responsibilities with data workers exclusively. Moreover, it could be helpful to investigate workers' dissent not as a hazard but as an asset that could help flag broader data quality issues, as Aroyo and Welty \cite{aroyo2015} have argued. A view into corporate work practices and market demands can offer a wider perspective to this line of research \cite{Posada2020a}.

Instead of technically correcting bias, this commentary is a call to “study up” machine learning data, that is, to interrogate the set of power relations that inscribe specific forms of knowledge in machine learning datasets. CSCW and HCI offer good examples of how different power conceptualizations can help broaden the study of socio-technical systems. For instance, scholars have drawn on feminist \cite{bardzell2011, muller2011, dignazio2020} and postcolonial \cite{philip2012, irani2013} theories to ask “Who” questions and make visible power dynamics in technoscientific discourses, highlighting their political nature. 

Our call also includes considering data workers as allies and assets in the quest for producing better and more just data, instead of portraying them as bias-carrying hazards. It means asking ourselves, “how is AI shifting power?” \cite{kalluri2020} rather than “how can worker biases be mitigated?”  Practitioners and researchers would do good by reflecting on power asymmetries that are inherent to creating data if the goal is accounting for “biased” data but, most importantly, for unjust socio-technical systems. Despite the abundant work (including several examples cited here) that has shown how power differentials shape data and data work, a number of investigations within our research community still direct their efforts towards mitigating biases in data work and crowdsourcing without considering the experiences and conditions of workers. Therefore, we insist on the need to foster interdisciplinary dialogue. Both lines of research \,---\, the study of power and the study of bias in ML data production \,---\, co-exist in parallel within CSCW and HCI. It is our hope that this commentary will prompt conversations that lead to more collaboration and, ultimately, to the advancement and broadening of this field of inquiry.

\subsection{How and Why Study Up Data?} 

We conclude by proposing a power-oriented research agenda to study ML data along three interrelated lines: 

\emph{First}, we propose conducting more qualitative and ethnographic research on data workers and data work production: Who are data workers? In what geographical and cultural contexts do they perform data work? What are the workflows, corporate infrastructures, and intra- and inter-organizational relationships in data production? How do these contexts affect data workers and dataset production? 
Retrieving data work settings can further make explicit the assumptions, norms, and values that inform and are inscribed in data work, allowing the “arenas of voice” \cite{star1999, Casilli2017a} and ethical considerations of workers \cite{Posada2021a} to emerge. In this sense, we argue that a deeper investigation into data workers and data work production cannot be achieved through mere quantitative measures and necessitates qualitative and exploratory research as well as the expertise of social scientists.  

\emph{Second}, we propose “shift[ing] the gaze upward” \cite{barabas2020} and studying the actors who outsource the creation of machine learning datasets: Who are data work requesters? What are their needs and wants? What rationale and priorities do they inscribe in data work tasks? What are the organizational forces driving them to produce and request data in specific ways? How do their needs and demands affect data workers’ labor conditions? 
Investigating the role of ML practitioners commissioning data-related tasks could help to explore the collaborative nature of data work and would see requesters as co-designers of data, and not as mere consumers. Here, too, it is important to look into the organizational settings in which the work of model developers is embedded. Drawing attention to data work requesters and their organizations can therefore reveal the service relationships, market logics, and the resulting power asymmetries that shape data work and, thereby, data.

\emph{Finally}, we propose expanding data documentating research and existing documentation frameworks: How can data documentation become sensitive to power relations and data production contexts? What would such a data documentation framework look like? How could organizations be incentivized to adopt such a documentation approach? How can we go beyond recognizing the power imbalances inscribed in data work and take action to bridge the power gap? Recognizing and investigating power relations are the initial steps to challenge them \cite{dignazio2020}. In this sense, a power-oriented data documentation framework can be one of the tools to render power \,---\, and its imbalances \,---\, visible in data work. In line with previous research \cite{gebru2020, madaio2020, miceli2021a}, we argue that documentation frameworks should be grounded on the needs of workers, be integrated into existing workflows and organizational infrastructure, and have the flexibility to accommodate specific work scenarios.

\begin{acks}
Funded by the German Federal Ministry of Education and Research (BMBF) – Nr 16DII113, the International Development Research Centre of Canada, and the Schwartz Reisman Institute for Technology and Society. We would like to acknowledge the data workers that have shared their knowledge and experience with us so that we could develop the ideas outlined in these pages. Special thanks to Alex Hanna, Bettina Berendt, and our anonymous reviewers for providing insightful comments that helped us strengthen our argument. 
\end{acks}

\bibliography{ref.bib}
\bibliographystyle{acm}

\end{document}